# The HAWC ultra-high-energy gamma-ray map with more than 5 years of data

**Kelly Malone,**[a] **José Andrés García González**[b] **and Patrick Harding**[c,*]

[a]*Space Science and Applications Group, Los Alamos National Laboratory, Los Alamos, NM, USA*
[b]*Tecnologico de Monterrey, Escuela de Ingeniería y Ciencias, Ave. Eugenio Garza Sada 2501, Monterrey, N.L., Mexico, 64849*
[c]*Physics Division, Los Alamos National Laboratory, Los Alamos, NM, USA*

 *E-mail:* kmalone@lanl.gov, anteus79@tec.mx, jpharding@lanl.gov

In 2020, the HAWC Collaboration presented the first catalog of gamma-ray sources emitting above 56 TeV and 100 TeV. With nine sources detected, this was the highest-energy source catalog to date. Here, we present the results of re-analysis of the old data, along with additional data acquired since then. We use a new version of the reconstruction software with better pointing accuracy and improved gamma/hadron separation. We now see more than 25 sources above 56 TeV, with most sources being located in the Galactic plane. The vast majority of these seem to be leptonic pulsar wind nebulae, but some have been shown to have hadronic emission. We will show spectra and discuss possible emission mechanisms of some of the most interesting sources, including the ones the HAWC Collaboration considers PeVatron candidates.

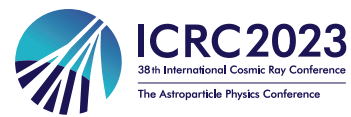

*Speaker

## 1. Previous observations of the UHE sky

Ulta-high-energy (UHE; > ~ 56 TeV) gamma-ray detections have long been of interest to the multi-messenger astrophysics community. Gamma-ray sources emitting above ~100 TeV could be "PeVatrons", or hadronic sources that emit Galactic cosmic rays to ~1 PeV, the knee of the cosmic ray spectrum.

Prior to 2019, there were no known gamma-ray sources known to emit in the ultra-high-energy regime. In that year, the paradigm changed. The Tibet-AS$\gamma$ and HAWC Collaborations both reported detections of the Crab Nebula emitting to UHE [1, 2]; this detection was shortly followed by the HAWC Collaboration's UHE catalog [3]. This catalog, the first of its kind, consisted of 9 sources emitting above 56 TeV, with three continuing above 100 TeV. It is interesting to note that all of the sources were within 0.5 degree of a pulsar, and that pulsar had an extremely high $\dot{E}$ (> $10^{36}$ erg/s) for 8 of the 9 sources. This provided hints that the UHE sky is more complex than previously thought, and that leptonic, not hadronic emission, may dominate at these energies.

The LHAASO collaboration also recently published a catalog of UHE sources [4]. This consisted of 12 sources emitting above 100 TeV, including a few sources that were new discoveries.

In this proceeding, we present a preliminary updated HAWC UHE catalog. Section 2 contains a description of HAWC, Section 3 contains a description of the catalog construction method, and Section 4 contains the results.

## 2. Description of HAWC

HAWC is an extensive air shower array located in the state of Puebla, Mexico. The detector consists of 300 tanks in the main array, 7.3 meters in diameter by 4.5 meters deep. HAWC is sensitive to gamma rays between ~300 GeV and several hundred TeV. There are also 345 outrigger tanks, which are smaller and located around the edges of the main array. They increase the sensitivity to the highest energies, but are not used in the analysis presented here. For more information about HAWC, see [5].

We use 2400 days of data, an increase of ~1361 days over the previous UHE catalog. We also use HAWC's 5th pass over the data, which has updated reconstruction algorithms. Most notably, this dataset contains better pointing accuracy and improved gamma/hadron separation algorithms. We now detect 28 sources emitting above 56 TeV, many of which cross the 100 TeV and 177 TeV thresholds at very high significance (> $5\sigma$).

## 3. Catalog construction method

First, maps above the three energy thresholds of interest (56 TeV, 100 TeV, and 177 TeV) are created using a likelihood framework [6]. The ground parameter energy estimator, described in [2], is used. The maps assume a power-law spectrum with a spectral index of -2.5 and a pivot energy of 7 TeV. Maps are created for a point source morphology, as well as extended disks with radii of 0.5, 1.0, 1.5, and 2.0 degrees. The only free parameter in the likelihood fit is the flux normalization.

For each pixel in each map, a test statistic (TS) is computed. This test statistic is defined as twice the likelihood ratio:

$$TS = 2ln(\frac{L_s}{L_n}) \quad (1)$$

where $L_s$ is the source hypothesis and $L_n$ is the null (background-only) hypothesis.

The UHE catalog is constructed using the same method used to construct the 3HWC catalog (HAWC's third all-energy catalog). More details can be found in [7], but a summary is provided below:

1. All hotspots in the point source map with a TS of > 25 are identified. In the event that multiple local maxima are found near each other, we separate the sources into primary and secondary sources. Primary sources are separated by nearby local maxima of higher significance by a valley of $\Delta(\sqrt{TS}) > 2$. Secondary sources (which have an asterisk next to their name on the source list) are separated from nearby sources by a TS valley of $1 < \Delta(\sqrt{TS}) < 2$.

2. We then run the same algorithm over the 0.5 degree extended map. To avoid finding sources that are in actuality two point sources smeared together when the extended maps are created, we only add a source to the list if it is more than 2 degrees from all known point sources.

3. We repeat step (2) for the remaining spatial morphologies (1.0, 1.5, and 2.0 degree disks).

Therefore, the search that a particular source is found in should be thought of as a lower limit on the emission. As an example, MGRO J1908+06, one of the brightest sources in the sky, is known to be extended [8], but is detected in the point source search here.

Additionally, very large extended sources may instead be miscategorized as several point sources. This is most evident in the region of Geminga and Monogem, which have angular extensions of several degrees wide.

Note that improvements to this catalog construction method are currently being developed by the HAWC collaboration. These will be addressed in future publications. A multi-source fitting method is under development and will be applied to the UHE dataset in the future. Additionally, many of the sources found here have had dedicated analyses, which take into account issues such as diffuse emission.

Due to the improvements to the method, the results should be considered preliminary. The 'xHWC' identifier is used in the results until the second UHE catalog is officially published.

## 4. The UHE sky

28 sources are detected with a TS of > 25 above 56 TeV. Five of those sources are secondary sources. Note that three of the sources are found in the Geminga/Monogem region, indicating that the algorithm does not perform well in regions of large, extended emission.

Most of the sources emitting above 56 TeV are in the inner Galactic plane - exceptions include the 3 Geminga-region sources, the Crab Nebula, and xHWC J5046+227*, believed to be the TeV halo close to the Crab Nebula[1]

Figures 1 through 3 show various sources emitting above 56 TeV. Table 1 shows the results of the catalog search.

---

[1]https://www.astronomerstelegram.org/?read=10941

| Name | Search | TS | RA [°] | Dec [°] | l [°] | b [°] | > 100 | > 177 |
|---|---|---|---|---|---|---|---|---|
| xHWC J0534+219 | PS | 584 | 83.67 | 21.98 | 184.60 | -5.77 | yes | |
| xHWC J0546+227* | 1.5° | 27.3 | 86.70 | 22.75 | 185.43 | -2.98 | | |
| xHWC J0633+179 | 0.5° | 36.1 | 98.31 | 17.97 | 194.88 | 4.22 | | |
| xHWC J0642+154* | 1.0° | 25.1 | 100.59 | 15.40 | 198.17 | 5.00 | | |
| xHWC J0657+140* | 2.0° | 25.9 | 104.37 | 14.09 | 200.99 | 7.69 | | |
| xHWC J1809-193 | PS | 91.7 | 272.42 | -19.31 | 11.07 | 0.08 | yes | |
| xHWC J1813-127 | PS | 25.5 | 273.30 | -12.71 | 17.27 | 2.52 | | |
| xHWC J1813-168* | PS | 35.4 | 273.47 | -16.88 | 13.69 | 0.37 | | |
| xHWC J1813-178 | PS | 95 | 273.30 | -17.82 | 12.78 | 0.07 | yes | |
| xHWC J1826-135 | PS | 306 | 276.50 | -13.59 | 17.97 | -0.65 | yes | yes |
| xHWC J1831-099 | PS | 41.3 | 277.87 | -9.97 | 21.79 | -0.15 | see Note1 | |
| xHWC J1833-091 | PS | 31.4 | 278.39 | -9.14 | 22.77 | -0.23 | see Note1 | |
| xHWC J1834-069 | PS | 27.4 | 278.61 | -6.92 | 24.84 | 0.60 | | |
| xHWC J1838-067 | PS | 49.3 | 279.54 | -6.73 | 25.43 | -0.13 | yes | yes |
| xHWC J1839-058 | PS | 117 | 279.89 | -5.83 | 26.39 | -0.02 | | |
| xHWC J1843-039 | PS | 44.1 | 280.99 | -3.92 | 28.59 | -0.12 | | |
| xHWC J1848+000 | PS | 146 | 282.22 | -0.04 | 32.61 | 0.55 | yes | yes |
| xHWC J1852+000 | PS | 73.1 | 283.01 | -0.04 | 32.97 | -0.15 | yes | |
| xHWC J1858+020 | PS | 84.8 | 284.50 | 2.05 | 35.51 | -0.53 | yes | |
| xHWC J1906+071* | PS | 31.7 | 286.57 | 7.11 | 40.94 | -0.04 | | |
| xHWC J1908+062 | PS | 137 | 287.10 | 6.28 | 40.45 | -0.89 | yes | yes |
| xHWC J1911+095 | 0.5° | 26.2 | 287.97 | 9.59 | 43.79 | -0.13 | yes | |
| xHWC J1928+180 | PS | 47.5 | 292.24 | 18.01 | 53.17 | 0.18 | yes | |
| xHWC J1928+189 | PS | 26 | 292.06 | 18.96 | 53.92 | 0.78 | | |
| xHWC J1959+287 | PS | 28 | 299.80 | 28.76 | 65.93 | -0.45 | | |
| xHWC J2019+367 | PS | 142 | 304.81 | 36.75 | 74.93 | 0.36 | yes | |
| xHWC J2031+413 | 0.5° | 49.7 | 307.79 | 41.36 | 80.03 | 1.13 | yes | yes |
| xHWC J2227+609 | PS | 26.9 | 336.79 | 60.91 | 106.27 | 2.79 | yes | yes |

**Table 1:** The source catalog. 'Search' refers to which extension search the source is found in, with 'PS' being the point source search. The TS and coordinates refer to the >56 TeV catalog search. Coordinates are given in both (right ascension, declination) and Galactic coordinates ($l, b$). Source positions may shift at higher energies. The last two columns denote whether a significant (TS > 25) detection is found above 100 TeV or above 177 TeV. Note1: xHWC J1831-095 is found above 100 TeV in the 1.0 extended souce search. This is 0.38 degrees from xHWC J1831-099 and 0.45 degrees away from xHWC J1833-091. It is unclear without further study which source the 100 TeV emission originates from.
.

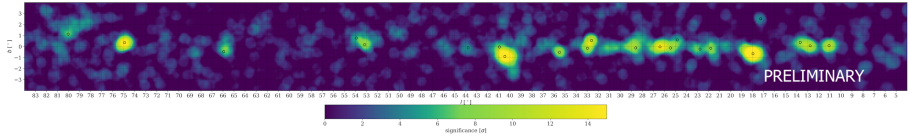

**Figure 1:** The Galactic plane above 56 TeV. As most sources in the Galactic plane have been shown to be extended in nature, the map assumes a 0.5 degree disk as the spatial morphology. Black circles denote the locations of sources emitting above 56 TeV in reconstructed energy.

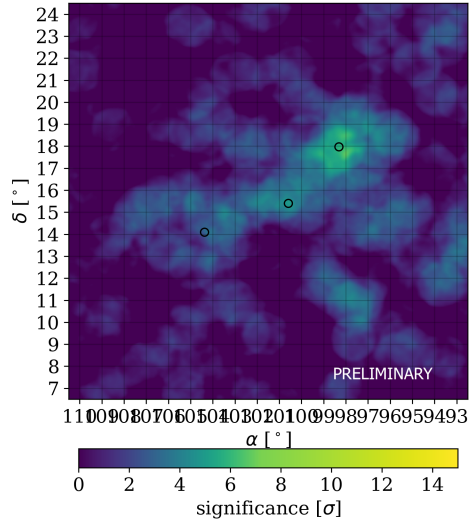

**Figure 2:** The Geminga and Monogem region above 56 TeV. The map assumes a 1 degree disk as the morphology. Note that three separate sources are found.

17 sources are detected above 100 TeV, and six sources are detected above 177 TeV. Figures 4 and 5 show the Galactic plane above these two energy thresholds. Due to space constraints, the coordinates of the >100 TeV and >177 TeV sources are not published here. Instead, Table 1 includes columns to denote if a > 56 TeV source continues emitting above these energy thresholds. In some situations, this association is hard to determine - since many of these sources are extended in nature, their coordinates can shift between the different energy threshold maps. This is most evident for xHWC J1831-095, which is detected above 100 TeV with the coordinates (right ascension, declination) = (277.87°, -9.59°). This is 0.38 degrees from xHWC J1831-099 and 0.45 degrees away from xHWC J1833-091, both detected in the > 56 TeV map. More study is needed to disentangle the different emission mechanisms in this region.

15 of the 17 sources emitting above 100 TeV appear to have a > 56 TeV counterpart. Of the two sources that do not have a counterpart, xHWC J2027+369 is a 2.0 degree source found in the Cygnus region with (right ascension, declination) = (307.18°, 41.66°). This region is known to have a large, extended emission region known as the Cygnus Cocoon [9], and may not have been found in the lower-energy map due to the selection criteria requiring sources be a certain distance apart. The second source, xHWC J2221-225 with (right ascension, declination) = (335.44°, -22.59°, is right at the detection threshold (TS = 25.6) and is not near any known sources in the TeVCat catalog. This source is likely a false positive. Characterizations of the false positive rate are currently underway.

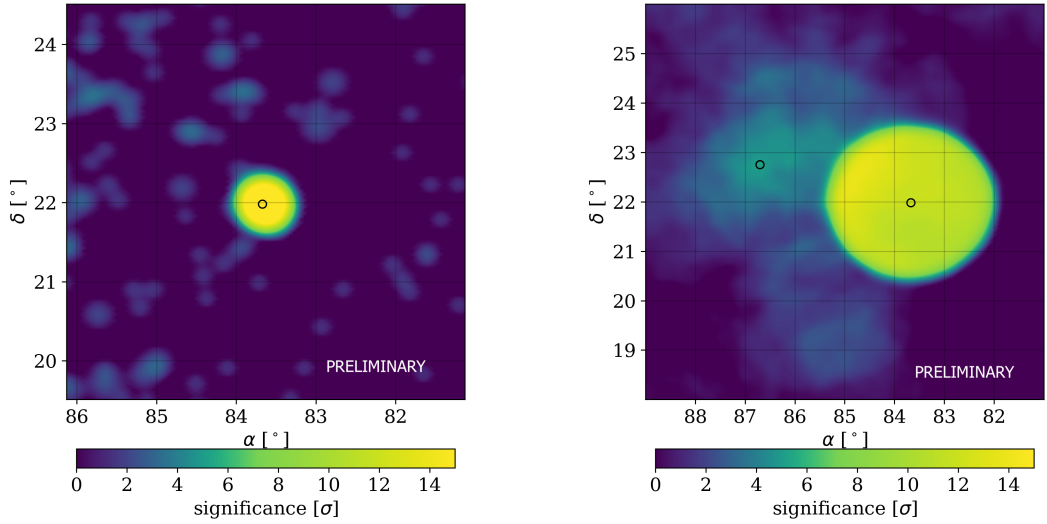

**Figure 3:** Left: The point source map above 56 TeV, showing the Crab Nebula emitting to very high significance. Right: The same region, but slightly zoomed out and smoothed by 1.5 degrees. A second source is clearly seen. This is believed to be the high-energy extension of a previously-reported TeV halo candidate.

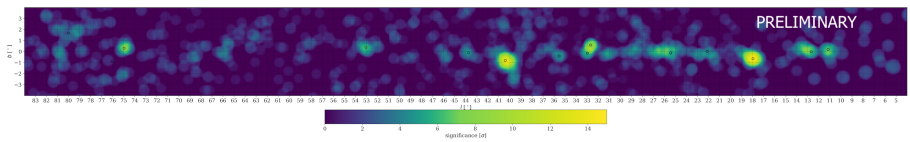

**Figure 4:** The Galactic plane above 100 TeV, assuming a 0.5 degree disk as the morphology. Black circles denote the locations of sources emitting above 100 TeV in reconstructed energy.

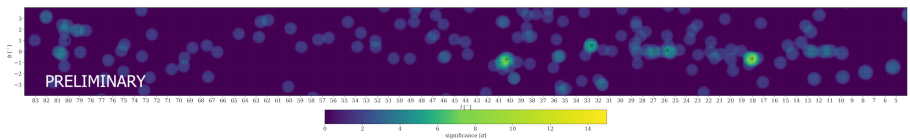

**Figure 5:** The Galactic plane above 177 TeV, assuming a 0.5 degree disk as the morphology. Black circles denote the locations of sources emitting above 177 TeV in reconstructed energy.

Likewise, five of the six sources detected above 177 TeV are also found in the >56 TeV map. These area also denoted in Table 1. The sixth source, xHWC J1609-093, is also far from the Galactic plane with (right ascension, declination) = (242.36°, -9.37°) and is right at the detection threshold (TS = 25.3).

## 5. Conclusions

We have presented here an update to the eHWC UHE catalog published in [3]. This updated catalog contains approximately three times as many sources as the original paper. A publication is forthcoming, which will include slight modifications to the catalog construction method as well as a quantification of the false detection rate.

## 6. Acknowledgements

We acknowledge the support from: the US National Science Foundation (NSF); the US Department of Energy Office of High-Energy Physics; the Laboratory Directed Research and Development (LDRD) program of Los Alamos National Laboratory; Consejo Nacional de Ciencia y Tecnología (CONACyT), México, grants 271051, 232656, 260378, 179588, 254964, 258865, 243290, 132197, A1-S-46288, A1-S-22784, CF-2023-I-645, cátedras 873, 1563, 341, 323, Red HAWC, México; DGAPA-UNAM grants IG101323, IN111716-3, IN111419, IA102019, IN106521, IN110621, IN110521 , IN102223; VIEP-BUAP; PIFI 2012, 2013, PROFOCIE 2014, 2015; the University of Wisconsin Alumni Research Foundation; the Institute of Geophysics, Planetary Physics, and Signatures at Los Alamos National Laboratory; Polish Science Centre grant, DEC-2017/27/B/ST9/02272; Coordinación de la Investigación Científica de la Universidad Michoacana; Royal Society - Newton Advanced Fellowship 180385; Generalitat Valenciana, grant CIDEGENT/2018/034; The Program Management Unit for Human Resources & Institutional Development, Research and Innovation, NXPO (grant number B16F630069); Coordinación General Académica e Innovación (CGAI-UdeG), PRODEP-SEP UDG-CA-499; Institute of Cosmic Ray Research (ICRR), University of Tokyo. H.F. acknowledges support by NASA under award number 80GSFC21M0002. We also acknowledge the significant contributions over many years of Stefan Westerhoff, Gaurang Yodh and Arnulfo Zepeda Dominguez, all deceased members of the HAWC collaboration. Thanks to Scott Delay, Luciano Díaz and Eduardo Murrieta for technical support.

# Full Author List: HAWC Collaboration

A. Albert[1], R. Alfaro[2], C. Alvarez[3], A. Andrés[4], J.C. Arteaga-Velázquez[5], D. Avila Rojas[2], H.A. Ayala Solares[6], R. Babu[7], E. Belmont-Moreno[2], T. Capistrán[4], Y. Cárcamo[26], A. Carramiñana[9], F. Carreón[4], U. Cotti[5], J. Cotzomi[26], S. Coutiño de León[10], E. De la Fuente[11], D. Depaoli[12], C. de León[5], R. Diaz Hernandez[9], J.C. Díaz-Vélez[11], B.L. Dingus[1], M. Durocher[1], M.A. DuVernois[10], K. Engel[8], C. Espinoza[2], K.L. Fan[8], K. Fang[10], N.I. Fraija[4], J.A. García-González[13], F. Garfias[4], H. Goksu[12], M.M. González[4], J.A. Goodman[8], S. Groetsch[7], J.P. Harding[1], S. Hernandez[2], I. Herzog[14], J. Hinton[12], D. Huang[7], F. Hueyotl-Zahuantitla[3], P. Hüntemeyer[7], A. Iriarte[4], V. Joshi[28], S. Kaufmann[15], D. Kieda[16], A. Lara[17], J. Lee[18], W.H. Lee[4], H. León Vargas[2], J. Linnemann[14], A.L. Longinotti[4], G. Luis-Raya[15], K. Malone[19], J. Martínez-Castro[20], J.A.J. Matthews[21], P. Miranda-Romagnoli[22], J. Montes[4], J.A. Morales-Soto[5], M. Mostafá[6], L. Nellen[23], M.U. Nisa[14], R. Noriega-Papaqui[22], L. Olivera-Nieto[12], N. Omodei[24], Y. Pérez Araujo[4], E.G. Pérez-Pérez[15], A. Pratts[2], C.D. Rho[25], D. Rosa-Gonzalez[9], E. Ruiz-Velasco[12], H. Salazar[26], D. Salazar-Gallegos[14], A. Sandoval[2], M. Schneider[8], G. Schwefer[12], J. Serna-Franco[2], A.J. Smith[8], Y. Son[18], R.W. Springer[16], O. Tibolla[15], K. Tollefson[14], I. Torres[9], R. Torres-Escobedo[27], R. Turner[7], F. Ureña-Mena[9], E. Varela[26], L. Villaseñor[26], X. Wang[7], I.J. Watson[18], F. Werner[12], K. Whitaker[6], E. Willox[8], H. Wu[10], H. Zhou[27]

[1]Physics Division, Los Alamos National Laboratory, Los Alamos, NM, USA, [2]Instituto de Física, Universidad Nacional Autónoma de México, Ciudad de México, México, [3]FCFM-MCTP, Universidad Autónoma de Chiapas, Tuxtla Gutiérrez, Chiapas, México, [4]Instituto de Astronomía, Universidad Nacional Autónoma de México, Ciudad de México, México, [5]Instituto de Física y Matemáticas, Universidad Michoacana de San Nicolás de Hidalgo, Morelia, Michoacán, México, [6]Department of Physics, Pennsylvania State University, University Park, PA, USA, [7]Department of Physics, Michigan Technological University, Houghton, MI, USA, [8]Department of Physics, University of Maryland, College Park, MD, USA, [9]Instituto Nacional de Astrofísica, Óptica y Electrónica, Tonantzintla, Puebla, México, [10]Department of Physics, University of Wisconsin-Madison, Madison, WI, USA, [11]CUCEI, CUCEA, Universidad de Guadalajara, Guadalajara, Jalisco, México, [12]Max-Planck Institute for Nuclear Physics, Heidelberg, Germany, [13]Tecnologico de Monterrey, Escuela de Ingeniería y Ciencias, Ave. Eugenio Garza Sada 2501, Monterrey, N.L., 64849, México, [14]Department of Physics and Astronomy, Michigan State University, East Lansing, MI, USA, [15]Universidad Politécnica de Pachuca, Pachuca, Hgo, México, [16]Department of Physics and Astronomy, University of Utah, Salt Lake City, UT, USA, [17]Instituto de Geofísica, Universidad Nacional Autónoma de México, Ciudad de México, México, [18]University of Seoul, Seoul, Rep. of Korea, [19]Space Science and Applications Group, Los Alamos National Laboratory, Los Alamos, NM USA [20]Centro de Investigación en Computación, Instituto Politécnico Nacional, Ciudad de México, México, [21]Department of Physics and Astronomy, University of New Mexico, Albuquerque, NM, USA, [22]Universidad Autónoma del Estado de Hidalgo, Pachuca, Hgo., México, [23]Instituto de Ciencias Nucleares, Universidad Nacional Autónoma de México, Ciudad de México, México, [24]Stanford University, Stanford, CA, USA, [25]Department of Physics, Sungkyunkwan University, Suwon, South Korea, [26]Facultad de Ciencias Físico Matemáticas, Benemérita Universidad Autónoma de Puebla, Puebla, México, [27]Tsung-Dao Lee Institute and School of Physics and Astronomy, Shanghai Jiao Tong University, Shanghai, China, [28]Erlangen Centre for Astroparticle Physics, Friedrich Alexander Universität, Erlangen, BY, Germany